\documentclass[twocolumn,prb,aps]{revtex4-1}
\usepackage{dcolumn}
\usepackage{amssymb}
\usepackage{bm}
\usepackage{graphicx}
\usepackage{float}
\usepackage{subfigure}
\usepackage{stackengine}
\usepackage{hyperref}
\hypersetup{colorlinks,linkcolor={red},citecolor={blue},urlcolor={blue}}
\newcommand{\e}[1]{e^{#1}} 
\newcommand{\avg}[1]{\left< #1 \right>} 
\newcommand{\abs}[1]{| #1 |} 

\begin{document}

\title{Entanglement conductance as a characterization of
  delocalized-localized phase transition in free fermion models}

\author{Mohammad Pouranvari}
\email{m.pouranvari@umz.ac.ir}
\affiliation {Department of Physics, Faculty of Basic Sciences,	University of Mazandaran, P. O. Box 47416-95447, Babolsar, Iran}

\author{Jahanfar Abouie}
\email{jahan@iasbs.ac.ir}
\affiliation {Department of Physics, Institute for Advanced Studies in Basic Sciences (IASBS), Zanjan 45137-66731, Iran}

\date{\today}

\begin{abstract}
  We study entanglement Hamiltonian (EH) associated with the reduced
  density matrix of free fermion models in delocalized-localized
  Anderson phase transition. We show numerically that the structure of
  the EH matrix differentiates the delocalized from the localized
  phase. In the delocalized phase, EH becomes a long-range Hamiltonian
  but is short-range in the localized phase, no matter what the
  configuration of the system's Hamiltonian is (whether it is long or
  short range). With this view, we introduce the entanglement
  conductance (EC), which quantifies how much EH is long-range and
  propose it as an alternative quantity to measure entanglement in the
  Anderson phase transition, by which we locate the phase transition
  point of some one-dimensional free fermion models; and also by
  applying the finite size method to the EC, we find three-dimensional
  Anderson phase transition critical disorder strength.
\end{abstract}

\maketitle

\section{Introduction}
Entanglement is a purely quantum concept: two particles that have
interacted in the past, never can be considered as two independent
particles\cite{PhysRev.47.777, schrodinger_1935}, rather they have to
be described as a unified entity. This concept is used in the quantum
information science as a physics resource with the applications in
quantum communication\cite{Duan2001, PhysRevLett.81.5932},
cryptography\cite{PhysRevLett.67.661, RevModPhys.74.145,
  PhysRevLett.68.3121}, teleportation\cite{PhysRevLett.80.869,
  Bouwmeester1997}, and computer sciences \cite{doi:10.1119/1.1463744,
  Kane1998, PhysRevA.52.R2493}. Later, condensed matter physicists
found this concept useful to characterize different phases, since
entanglement indirectly measures the amount of correlation in the
system. To quantify entanglement in a system, people have used
different measures. Entanglement Entropy (EE) is the most famous
candidate in a pure ground sate of a system; although there are other
alternative measures available for the ground state, and for excited
state\cite{RevModPhys.81.865} where the system is in a mixed
state. People in addition developed methods to measure entanglement in
experiment.\cite{Sackett2000, RevModPhys.73.565, 2018arXiv180804471C}

Among many other applications, EE can be specially beneficial for a
delocalized-localized phase transition, where state of the system
changes as disorder in the system is varied. In the delocalized phase,
the correlation in the system is larger than in the localized phase
and thus we see larger
EE.\cite{PhysRevLett.99.126801} Delocalized-localized phase transition
is manifested in lattice systems by Anderson
model\cite{PhysRev.109.1492} which is a tight binding model with
constant tunneling amplitude and random on-site energies. With
uncorrelated disorder\cite{RevModPhys.80.1355}, we know that the
system is localized with any infinitesimal amount of disorder in one
and two dimensions, and thus there is no Anderson phase
transition. While in three dimensions $3D$, at a critical disorder,
scattering of fermions by impurities becomes completely destructive
and state of the system becomes
localized.\cite{2006AcPSl..56..561M} However, with correlated
disorder, systems in $1D$ and $2D$ can also exhibit Anderson phase
transition\cite{RevModPhys.80.1355}, some of which are used in this
paper to verify our idea.

Anderson phase transition happens at zero temperature where
fluctuations has quantum nature only. It is among the class of second
order phase transition, where observables of the system at the phase
transition point become length-scale independent and by finite size
scaling one finds the phase transition point and the corresponding
universal critical exponents.\cite{Slevin_2014}

Aside from the EE, it is also found the entanglement
\emph{spectrum}\cite{PhysRevLett.101.010504, PhysRevB.95.115122,
  0295-5075-119-5-57003, PhysRevB.81.064439, PhysRevA.78.032329,
  PhysRevLett.108.196402} and also the \emph{eigenmodes} of the
entanglement Hamiltonian\cite{PhysRevB.89.115104, PhysRevB.88.075123, PhysRevB.92.245134} are useful to
characterize different phases. What is left, is to look at the
\emph{entanglement Hamiltonian} (EH) matrix to see what information
about the system we can catch. EH of a subsystem has been studied from
another perspectives. In a study\cite{1751-8121-50-28-284003}, the
explicit expression for the EH matrix elements in the ground state of
free fermion models has been reported. In another
study\cite{PhysRevB.99.235109}, operator form of the EH is constructed
based on one entangled mode of the reduced density matrix. People also
found that at the extreme limit of strong coupling between two chosen
subsystems, EH of a subsystem and its Hamiltonian are proportional.
\cite{0295-5075-96-5-50006, Moradi_2016} On the other hand, for a
non-zero temperature, in a highly excited state, reduced density of a
subsystem becomes the thermal density and correspondingly, EH of
subsystem relates to the subsystem's
Hamiltonian.\cite{PhysRevX.8.021026} In this paper, we emphasize on
the fact that structure of the ground state EH of a chosen subsystem
possesses physical information and thus useful to distinguish
different phases of the system. More specifically, we show that,
subsystem EH made by ground state of the whole system in a free
fermion model, has distinguishable configurations in delocalized and
localized phases of Anderson phase transition: EH is long-range in
delocalized phase and short-range in localized phase. In addition, to
quantify EH configuration, we use the notion of conductance of
entanglement Hamiltonian. By using EH conductance as an indicator, we
distinguish localized and delocalized phases, and also we locate exactly the
Anderson phase transition point.

This paper is structured as follows: in Section \ref{model+method} the
$1D$ models and the $3D$ Anderson model that we use in this paper are
explained; we also shortly explain how to obtain the EH for the ground
state. In Section \ref{EHindloc+loc}, EH structures in both
delocalized and localized phases are studied and contrasted. Then we
introduce EH conductance in section \ref{EHcond} as an indicator of
delocalized and localized phases. The conclusion and some suggestions
for future works are presented in section \ref{conc}.

\section{Models and Method}\label{model+method}

We consider $1D$ free fermion models and also the $3D$ Anderson
model. In the following, we introduce these models and review their
delocalized-localized phase transitions that has been proved
analytically and numerically before.

The first model we consider, is the generalized Aubry-Andre (gAA) model. It is a
$1D$ tight binding model with constant nearest-neighbor hopping
amplitude $t$:
\begin{equation}\label{Hccdagger}
 \mathcal{H}= \sum_{<ij>}t (c_i^{\dagger} c_{j} + c_{j}^{\dagger} c_{i}) +
  \sum_i \phi_i c_i^{\dagger} c_{i}
\end{equation}
where $<ij>$ stands for nearest-neighbor hopping and the on-site
energies $\phi_i$ have an incommensurate periodicity with respect to
lattice constant (set to $1$ here):
\begin{equation}\label {gAA}
  \phi_i = 2 \lambda \frac{\cos{(2\pi i b)}}{1-\alpha\cos{(2\pi i b)}},
\end{equation}
which $b = \frac{1+\sqrt{5}}{2}$ is the golden ratio (we set $t=-1$ in
our calculations). This model is neither completely periodic nor
completely random and as illustrated in Ref.
[\onlinecite{PhysRevLett.114.146601}], it has the separating mobility
edges of localized and delocalized states at the energy:
\begin{equation}\label{me}
   E_{\text{mobility edge}} = 2 sgn(\lambda)\frac{\abs{t}-\abs{\lambda}}{\alpha}.
\end{equation}

Note that this model has no randomness and delocalized phase happens
by the incommensurate periodicity of on-site energy. The special case
of Eq. (\ref{gAA}) with $\alpha=0$ is the Aubry-Andre (AA)
model\cite{aubry1980analyticity} with delocalized ($\lambda < 1$) and
localized ($\lambda > 1$) phases. AA model has no mobility edges,
i.e. all states become localized in the localized phase.

Another model is the power-law random banded matrix model
(PRBM)\cite{PhysRevE.54.3221} that is a long-range hopping model with
the following Hamiltonian:
\begin{equation}\label{Hhij}
  \mathcal{H}= \sum_{ij}h_{ij} c_i^{\dagger} c_{j},
\end{equation}
in which matrix elements $h_{ij}$ are randomly Gaussian distributed
numbers, with zero mean and the following variance (if we use the
periodic boundary condition):
\begin{equation}
  \avg {\abs{h_{ij}}^2} = \left[{1+\left(\frac{\sin{\pi
(i-j)/N}}{b \pi /N}\right)^{2\alpha}}\right]^{-1},
\end{equation}
where $N$ is the system size and we set $b=1$. In the limiting case of $\alpha \gg 1$, the
variance $\avg {\abs{h_{ij}}^2}$ approaches zero for the next nearest
neighbor couplings and further, and thus the Hamiltonian of the system
will be a Hamiltonian with short-range couplings. On the other hand,
when $\alpha \ll 1$, $\avg {\abs{h_{ij}}^2}$ approaches to $1/2$ for
all couplings, thus yield to a long-range Hamiltonian with all
couplings to be non-zero. Therefore, the system goes through Anderson
phase transition at $\alpha=1$, it is in delocalized phase for
$\alpha<1$ and localized for $\alpha>1$.\cite{PhysRevE.54.3221} This
model is distinguished and important since different models can be
simulated by modifying the $b$ parameter.\cite{PhysRevLett.56.290,
  PhysRevLett.76.2386, ALTSHULER1997487, PhysRevB.56.3742}

One another model we consider is the power-law random bond disordered
Anderson model (PRBA)\cite{PhysRevB.69.165117} which is a $1D$ model
with the Hamiltonian of Eq. (\ref{Hhij}), where on-site energies are
zero, and long-range hopping amplitudes are
\begin{equation}
  h_{ij}=\frac{w_{ij}}{|i-j|^{\alpha}},
\end{equation}
where $w_{i,j}$'s are uniformly random numbers distributed between $-1$ and
$1$. When $\alpha\ll 1$, hopping amplitude becomes slow decaying, and
the Hamiltonian is long-range. On the other hand, for $\alpha \gg 1$,
hopping amplitude goes very fast to zero and we have a short-range
Hamiltonian. Therefore, there is a phase transition at $\alpha=1$
between delocalized state ($\alpha<1$ with long-range hopping
amplitudes) and localized state ($\alpha>1$, with short-range hopping
amplitudes).

Finally, we also consider the three dimensional Anderson model (the
$3D$ version of Eq. (\ref{Hccdagger})) with constant nearest-neighbor
hopping amplitudes, $t=-1$, and randomly distributed on-site
energies. We use Gaussian distribution with mean zero and variance $W$
where the Anderson phase transition happens at
$W \approx 6$\cite{Slevin_2014}, the system is in delocalized
phase for $W<6$ and localized for $W>6$.

We note that in gAA and Anderson $3D$ models, the structure of the
Hamiltonian matrix is the same in delocalized and localized phases (it
is always a short-range Hamiltonian: only the nearest neighbor hopping
amplitude is non-zero). However, in the PRBM and PRBA models, the
structure of the Hamiltonian matrix is different in the delocalized
and localized phases: in the localized phase it is a short-range and
in the delocalized phase it is a long-range.

\subsection{Entanglement Hamiltonian (EH) for free fermion models}
Next, we explain the procedure to obtain Entanglement Hamiltonian for
free fermion models. One usually divides the system into two parts in
real space, subsystem $A$ form site $1$ to $N_A$ and the rest of the
system as subsystem $B$. Other type of partition has also been
used.\cite{PhysRevLett.110.046806, PhysRevB.90.104204,
  PhysRevB.88.115114} Then, EE is obtained by calculating the von
Neumann entropy of the reduced density matrix (RDM) of a chosen
subsystem. That is EE = $-tr \rho_A \log \rho_A$, where $\rho_A$ is
the RDM of subsystem $A$ computing by tracing over degrees of
freedom of subsystem $B$. Since the RDM is a positive definite
operator, we can write it as:
\begin{equation}
  \rho_A = \e{-\mathcal{H}^{ent}},
\end{equation}
where $\mathcal{H}^{ent}$ is called \emph{entanglement Hamiltonian}
(EH). For free fermion models (that we consider in this paper) EH is a
free fermion Hamiltonian:
\begin{equation}\label{Henthij}
  \mathcal{H}^{ent}= \sum_{ij}^{N_A}H^{ent}_{ij} c_i^{\dagger} c_{j}.
\end{equation}

To obtain the EH numerically in the free fermion models, one first
calculates the correlation matrix for the chosen subsystem:
\begin{equation}
C^A_{ij} = \langle c_i^{\dagger} c_{j} \rangle, \quad i,j =
1,\cdots, N_A.
\end{equation}
In free fermion models that we consider in this paper, we can
calculate the correlation matrix based on the eigenvectors of the
Hamiltonian, $V$:
\begin{equation}
C^A_{ij} = \sum_{k=1}^{N_{\rm F}} V_{ik} V^{\dagger}_{kj},
\end{equation}
where $N_{\rm F}$ is the number of fermions. By setting the Fermi energy
$E_{\rm F}$, we can obtain the number of fermions $N_{\rm F}$: we fill up the
energy levels by fermions until we reach the $E_{\rm F}$. In this paper we
set $E_{\rm F}=0$ and for each model and sample we calculate numerically the
number of fermions (only for gAA model we change Fermi energy from its
lowest value to its highest value and then obtain the number of
fermions accordingly).

Correlation matrix is related to $H^{ent}$
as\cite{0305-4470-36-14-101, PhysRevB.69.075111}:
\begin{equation}\label{HrC}
  H^{ent} = \ln \frac{1 -C^A}{C^A}.
\end{equation}
Diagonalizing the correlation matrix and finding its eigenvalues and
eigenvectors, we obtain the EH matrix.

\section{Entanglement Hamiltonian in delocalized-localized
  phases}\label{EHindloc+loc}
In the following we give a picture of the matrix elements of the EH in
localized and delocalized phases.  Matrix elements of the EH based on
Eq. (\ref{HrC}) are:
\begin{equation}\label{hentijU}
  H_{ij}^{ent} = \sum_{\ell=1}^{N_A} U_{i \ell} \ln
  \frac{1-\zeta_{\ell}}{\zeta_{\ell}} U^{\ast}_{j\ell}, 
\end{equation}
in which $\{\zeta_{\ell}\}$ and $U$ are the eigenvalues, and the
unitary matrix to diagonalize the correlation matrix, respectively.

\begin{figure}
  \centering
  \includegraphics[width=0.5\textwidth]{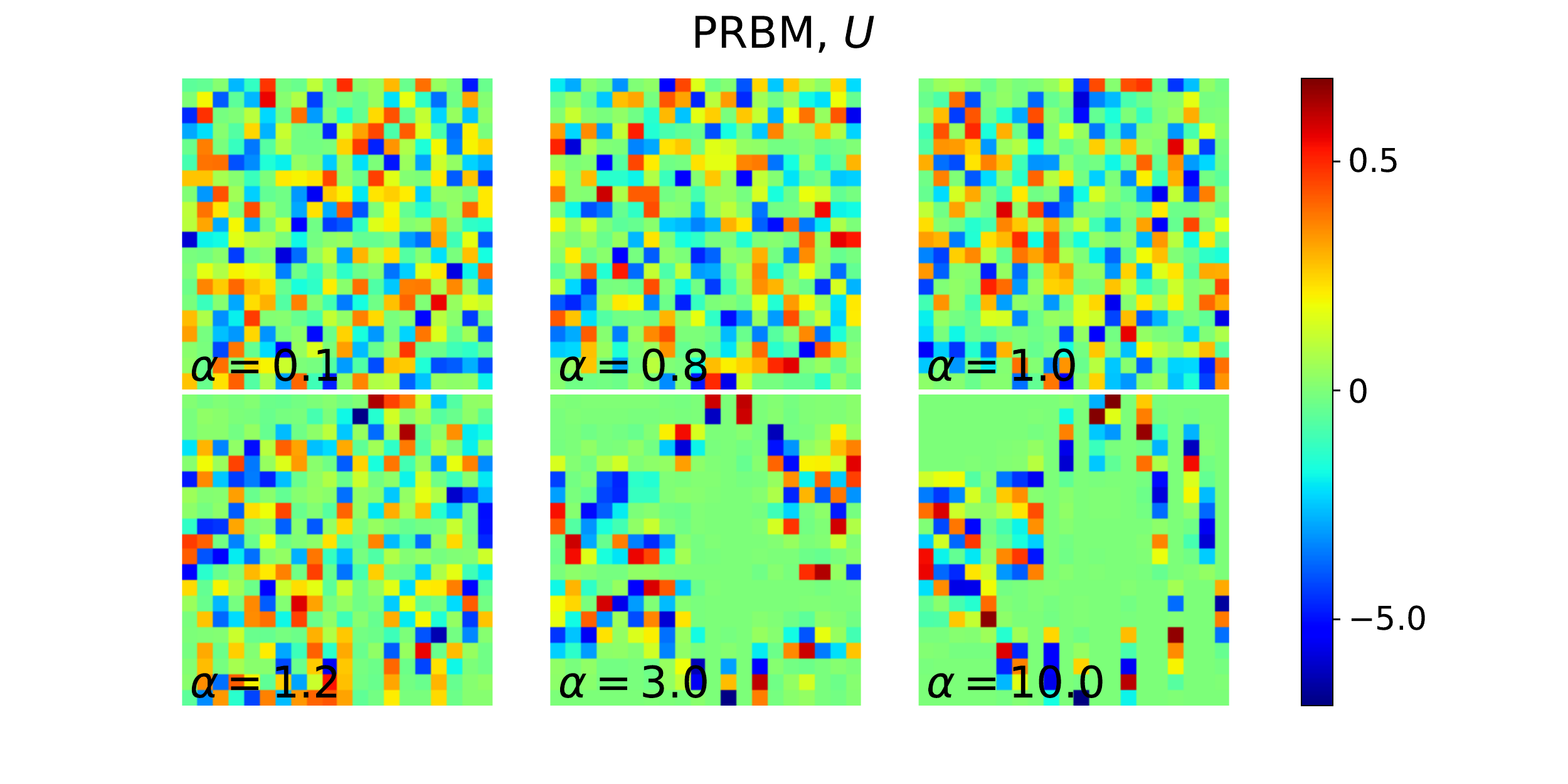}
  \caption{The eigenvectors of EH, i.e. $U$ matrix elements plotted
    for PRBM model with $N=40$ sites and $N_A=20$, as we increase
    $\alpha$ and go from delocalized phase ($\alpha<1$) to localized
    phase ($\alpha>1$). Each column is a normalized eigenvector of
    EH. Deep in the delocalized phase, it is completely extended over
    all sites, and it becomes localized only over a few sites in
    localized phase. Plots from top-left to bottom-right correspond to
    $\alpha=0.1$, 0.8, 1.0, 1.2, 3.0, and 10.0, respectively.}
  \label{fig:PRBM_Uent_N_40}
\end{figure}

One special eigenvector of the $U$ matrix, which corresponds to the
$\zeta$ closes to $1/2$, has proven to be localized (extended) in the
localized (delocalized) phase.\cite{PhysRevB.89.115104,
  PhysRevB.88.075123, PhysRevB.97.125116,
  doi:10.1142/S0217732318500852} Here, we show that all eigenvectors
of EH has this property. To verify it numerically, we plot the $U$
matrix composed of the EH eigenvectors for the PRBM model in
Fig. \ref{fig:PRBM_Uent_N_40}; each normalized eigenvector of EH
(which is a column in the $U$ matrix) is extended in the delocalized
phase and it has only a few non-zero values in the localized
phase. Same results obtained for other models we considered in this
paper (not shown). Thus, for each eigenvector, a localization length
can be defined over which the eigenvector is extended and outside of
which it vanishes.

\begin{figure}
  \centering
  \includegraphics[width=0.5\textwidth]{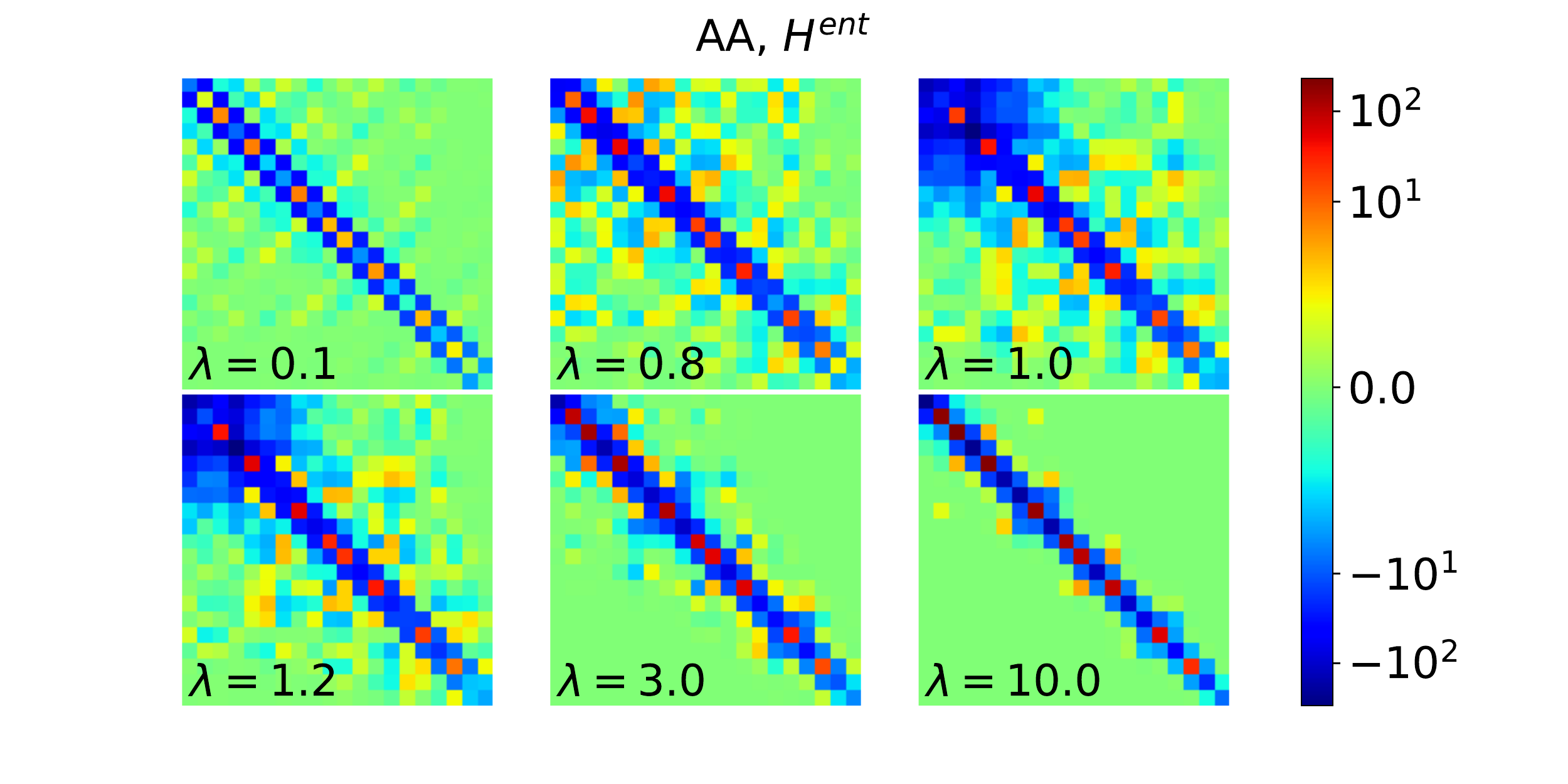}
  \caption{The EH matrix elements plotted for AA model of $N=40$
    sites, and $N_A=20$, as we increase $\lambda$ and go from
    delocalized phase ($\lambda<1$) to localized phase
    ($\lambda>1$). We set Fermi energy $E_{\rm F}=0$.}
  \label{fig:AA_Hent_N_40}
\end{figure}

\begin{figure}
  \centering
  \includegraphics[width=0.5\textwidth]{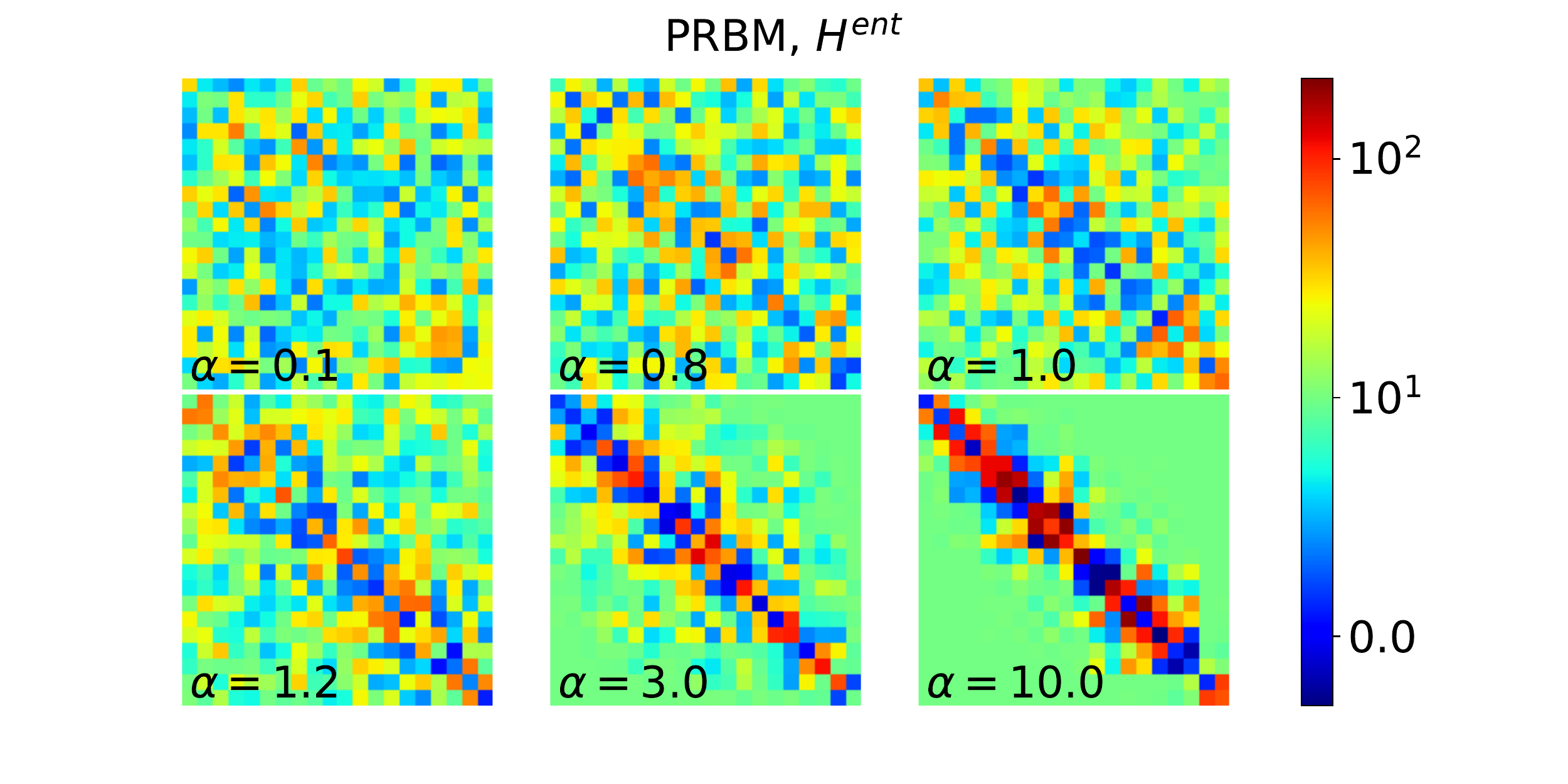}
  \caption{The EH matrix elements of PRBM model with $N=40$ sites and
    $N_A=20$, as we increase $\alpha$ and go from delocalized phase
    ($\alpha<1$) to localized phase ($\alpha>1$). We set Fermi energy
    $E_{\rm F}=0$. }
  \label{fig:PRBM_Hent_N_40}
\end{figure}

\begin{figure}
  \centering
  \includegraphics[width=0.5\textwidth]{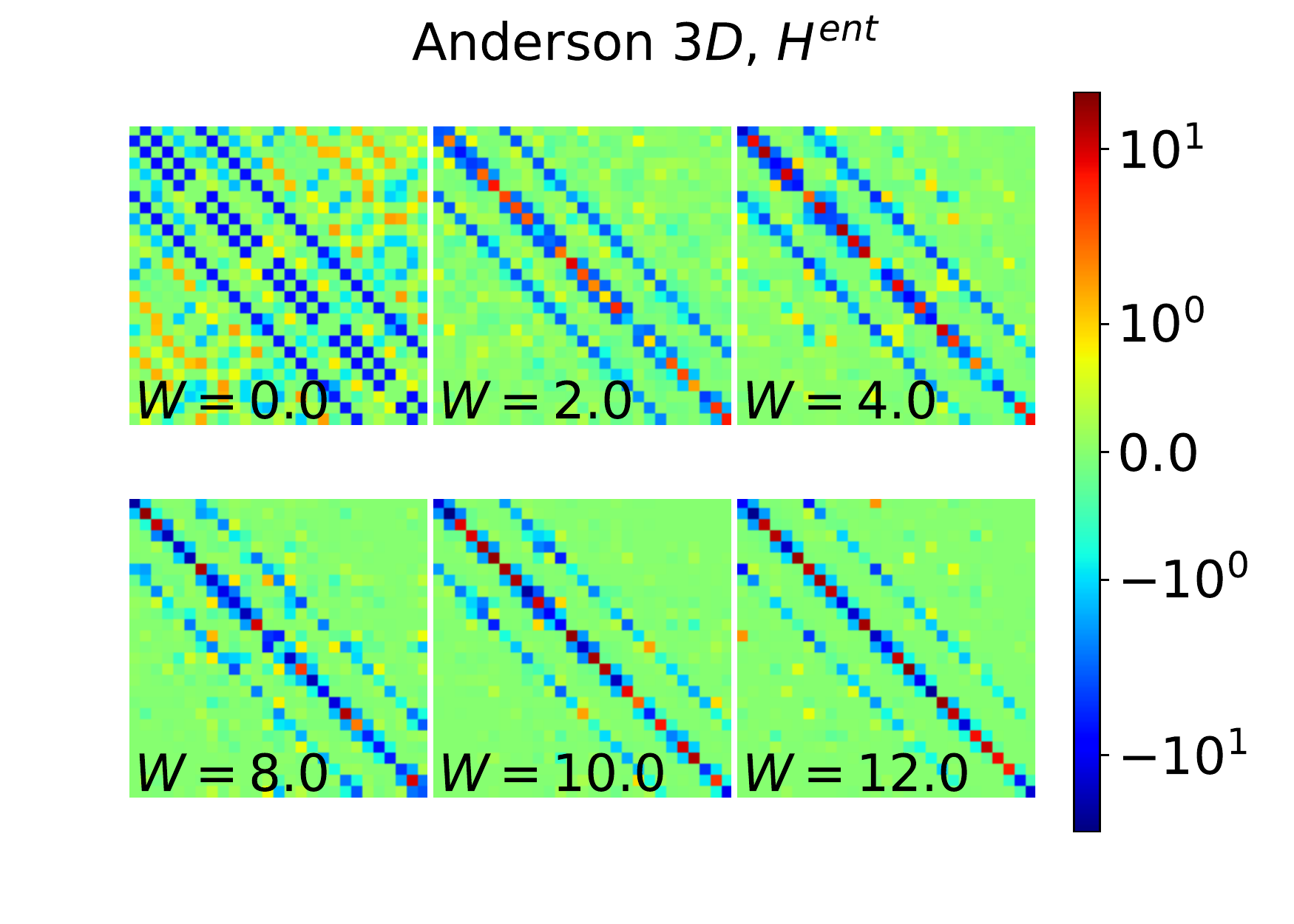}
  \caption{The EH matrix elements plotted for Anderson $3D$ model,
    with $N=6\times 6 \times 6$ and $N_A=N/2$, as we increase the
    disorder strength $W$ and go from delocalized phase ($W<6$) to
    localized phase ($W>6$). We set Fermi energy $E_{\rm F}=0$. We plot only
    one quarter of the EH. We note that Anderson $3D$ model has
    mobility edges at both tail of the spectrum, separating the
    delocalized and localized phases. Three upper plots correspond to
    the case when we are in the delocalized phase and three lower case
    correspond to the localized phase.}
  \label{fig:Anderson3D_Hent_multi}
\end{figure}

Now, having in mind this localization properties of $U$, we can
analyze the EH matrix elements. To obtain $H^{ent}_{i,j}$, as we go
from diagonal elements outward, i.e. for $H^{ent}_{i,i+n}$ elements,
as we increase $n$, we multiply $U_{i\ell}$ and $U_{i+n,\ell}$ when we
sum over $\ell$ in Eq. (\ref{hentijU}). For the $\ell$th eigenvector,
if both $i$ and $i+n$ elements are inside the localization length, we
will have a non-zero result; but as soon as one of these elements
falls outside of localization length, we will have a vanishing
result. In the localized phase where columns of $U$ matrix become
localized, then we get faster vanishing results for $H^{ent}_{ij}$ as
we go off-digonal. On the other hand, in the delocalized phase with
extended EH eigenvectors, localization length increases and we have
more non-zero elements for $H^{ent}_{ij}$. Therefore, in the localized
phase we expect to have a short-range hopping matrix for the EH with
only a few non-zero off-diagonal elements, while in the extended phase
the EH becomes long-range. EH of AA, PRBM and Anderson $3D$ models are
plotted in Figs.  \ref{fig:AA_Hent_N_40}, \ref{fig:PRBM_Hent_N_40},
and \ref{fig:Anderson3D_Hent_multi}, respectively, in delocalized and
localized phases. As we can see, the EH is a long-range Hamiltonian
(with many non-zero off-diagonal elements) in the delocalized phase,
and as we approach to the localized phase, it becomes short-range. In
the extreme situation, deep in the localized phase, diagonal elements
become much larger and the off diagonal elements become zero; on the
other hand, deep in delocalized phase, diagonal elements become
zero. Similar results are obtained for PRBA and gAA models (not shown
here).

This observation is true either for systems with Hamiltonians that are
short-range in both localized and delocalized phases (AA, and Anderson
models), or for system that its Hamiltonian is short-range in
localized and long-range in delocalized phase (PRBM, and PRBA
models). Thus, no matter the structure of the Hamiltonian of the
system is, the entanglement Hamiltonian is long-range in delocalized
phase and short-range in localized phase.

\section{Entanglement Conductance (EC)}\label{EHcond}
\begin{figure*}
  \centering
  \begin{subfigure}{}%
    \includegraphics[width=0.32\textwidth]{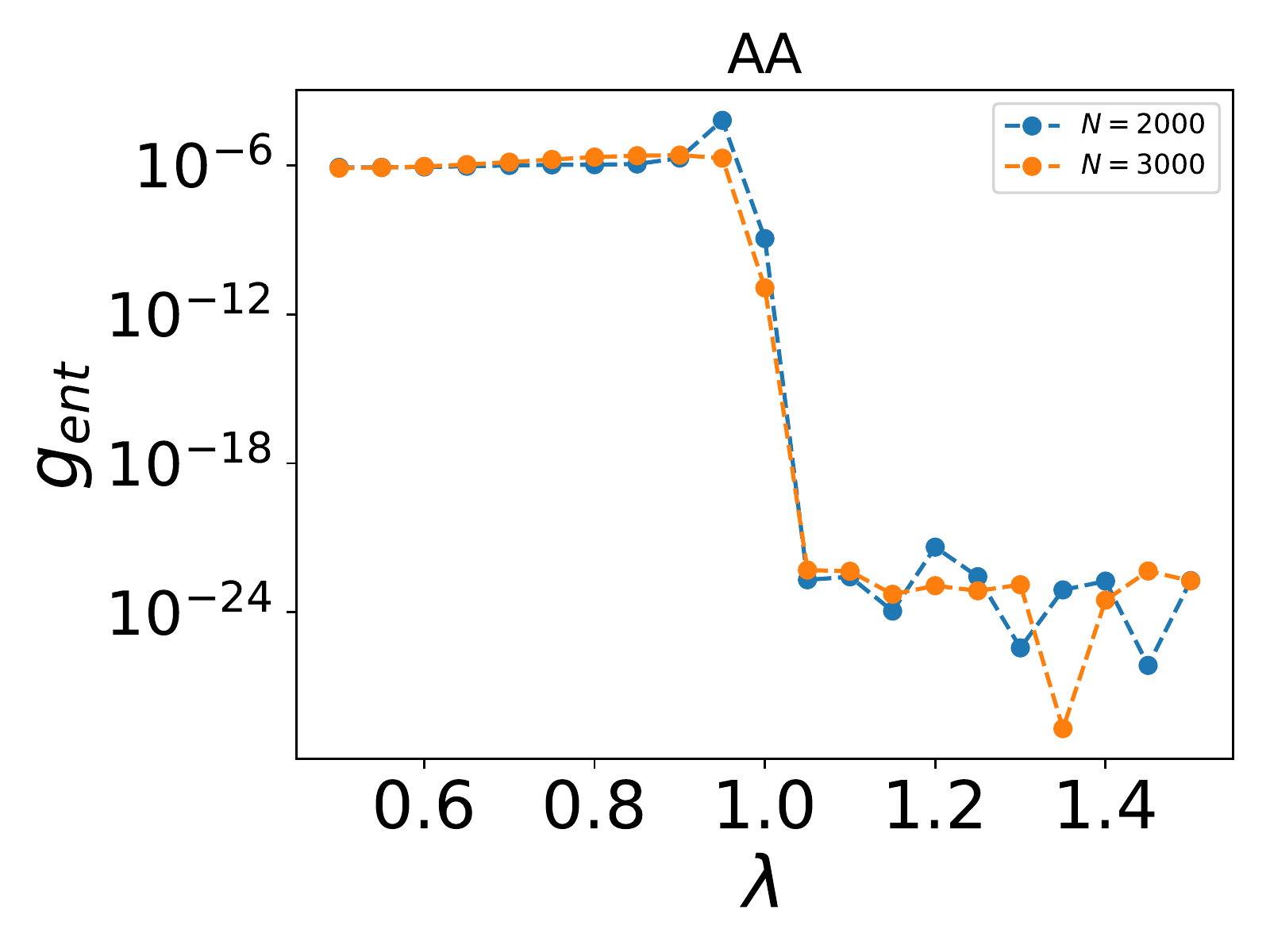}
  \end{subfigure}%
  ~%
  \begin{subfigure}{}%
    \includegraphics[width=0.32\textwidth]{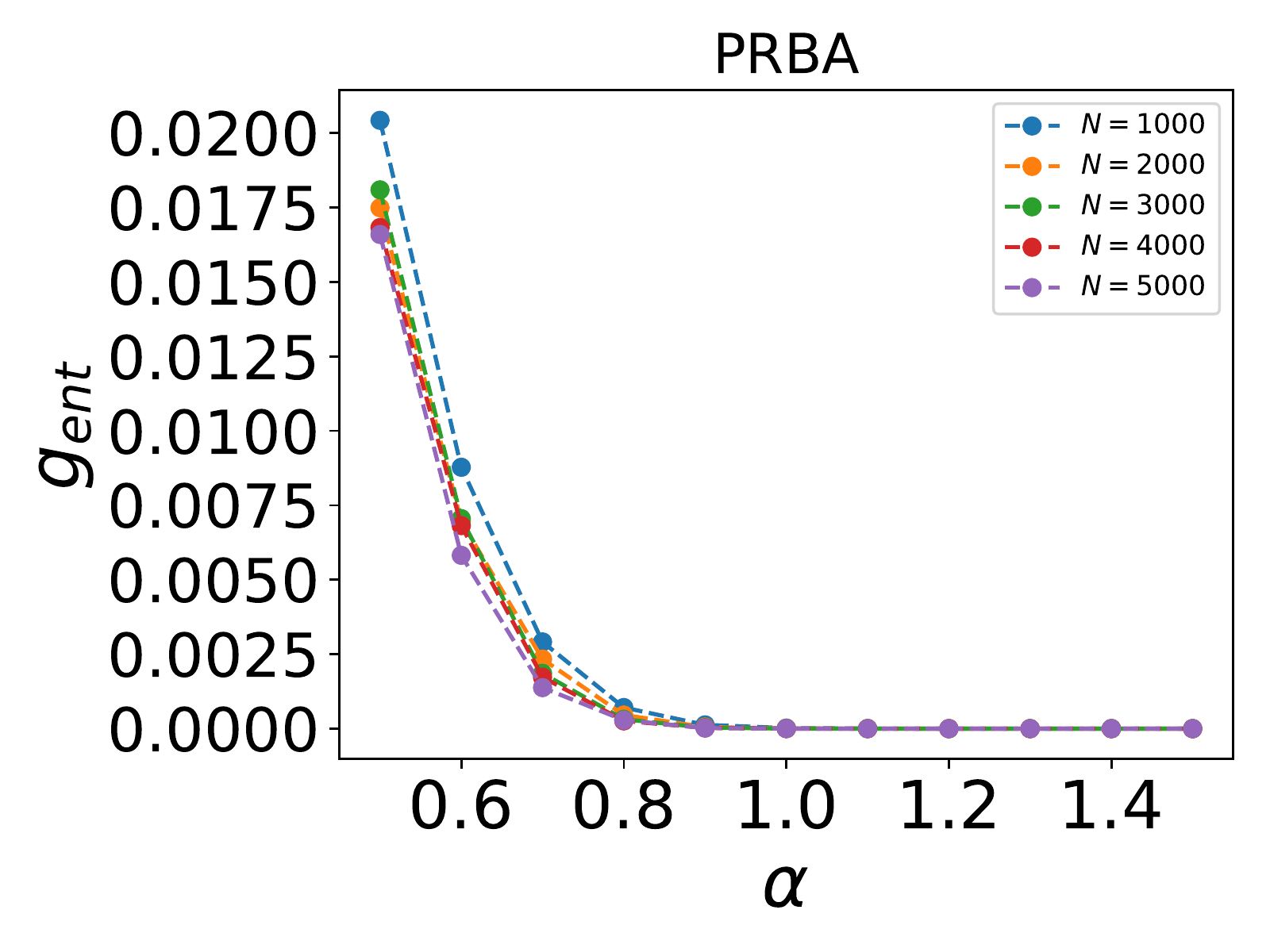}
  \end{subfigure}
    ~%
  \begin{subfigure}{}%
    \includegraphics[width=0.32\textwidth]{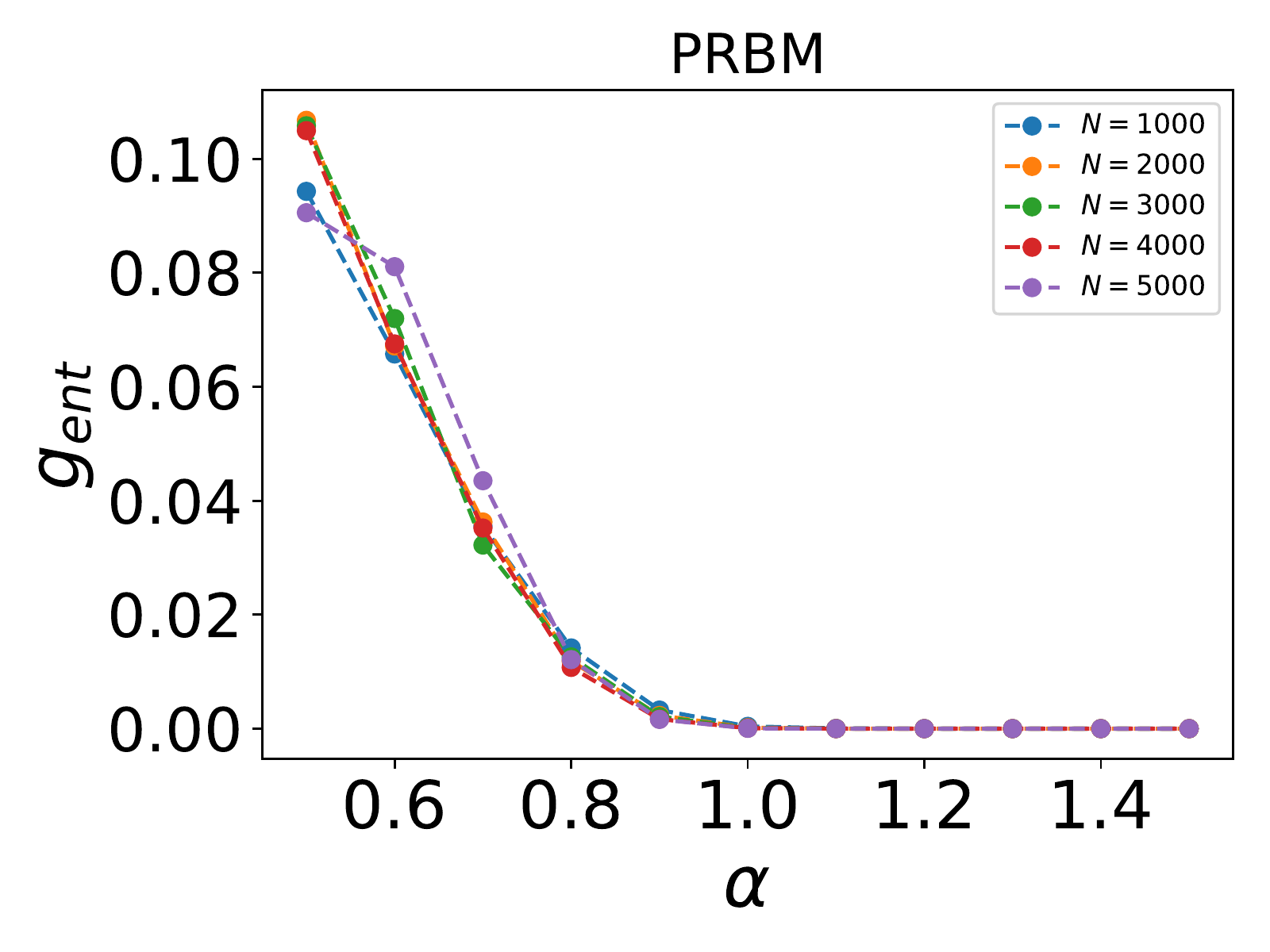}
  \end{subfigure}
  \caption{The EC ($g_{ent}$) for the AA model versus $\lambda$
    (left), for the PRBA model versus $\alpha$ (middle), and for the
    PRBM model versus $\alpha$ (right).  In three cases $g_{ent}$ goes
    to zero in the localized phase and it determines the phase
    transition point, exactly. In AA model we do not have randomness,
    so one sample is considered. In PRBA, the number of sample systems
    varies between $10^5$ for small $N$ and $10^3$ for the largest
    $N$.  In PRBM, the number of sample systems varies between $10^4$
    for small $N$ and $10^3$ for the largest
    $N$. \label{fig:g_AA_PRBA_PRBM}}
\end{figure*}

\begin{figure*}
  \centering
  \begin{subfigure}{}%
    \includegraphics[width=0.45\textwidth]{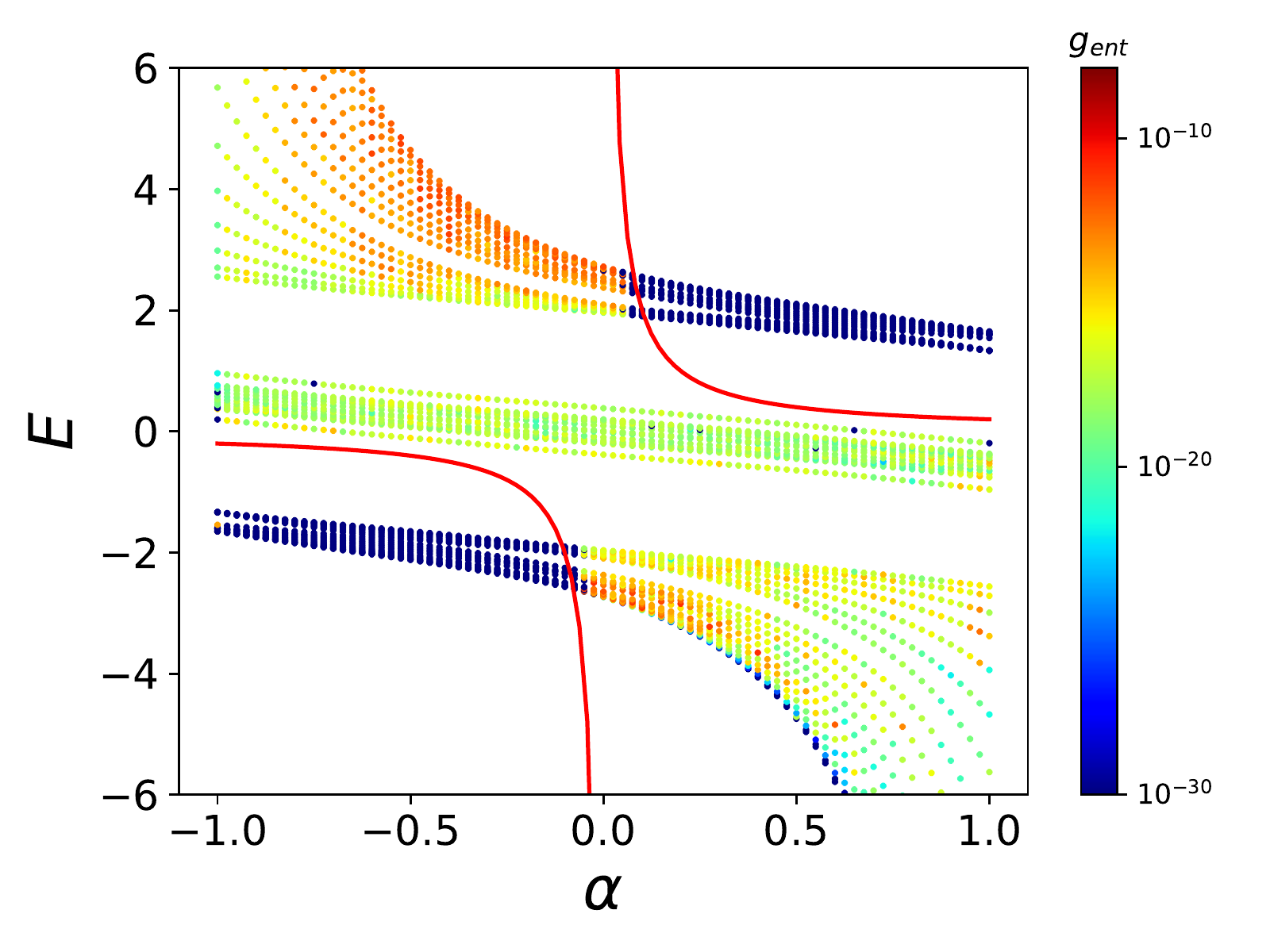}
  \end{subfigure}%
  ~%
  \begin{subfigure}{}%
    \includegraphics[width=0.45\textwidth]{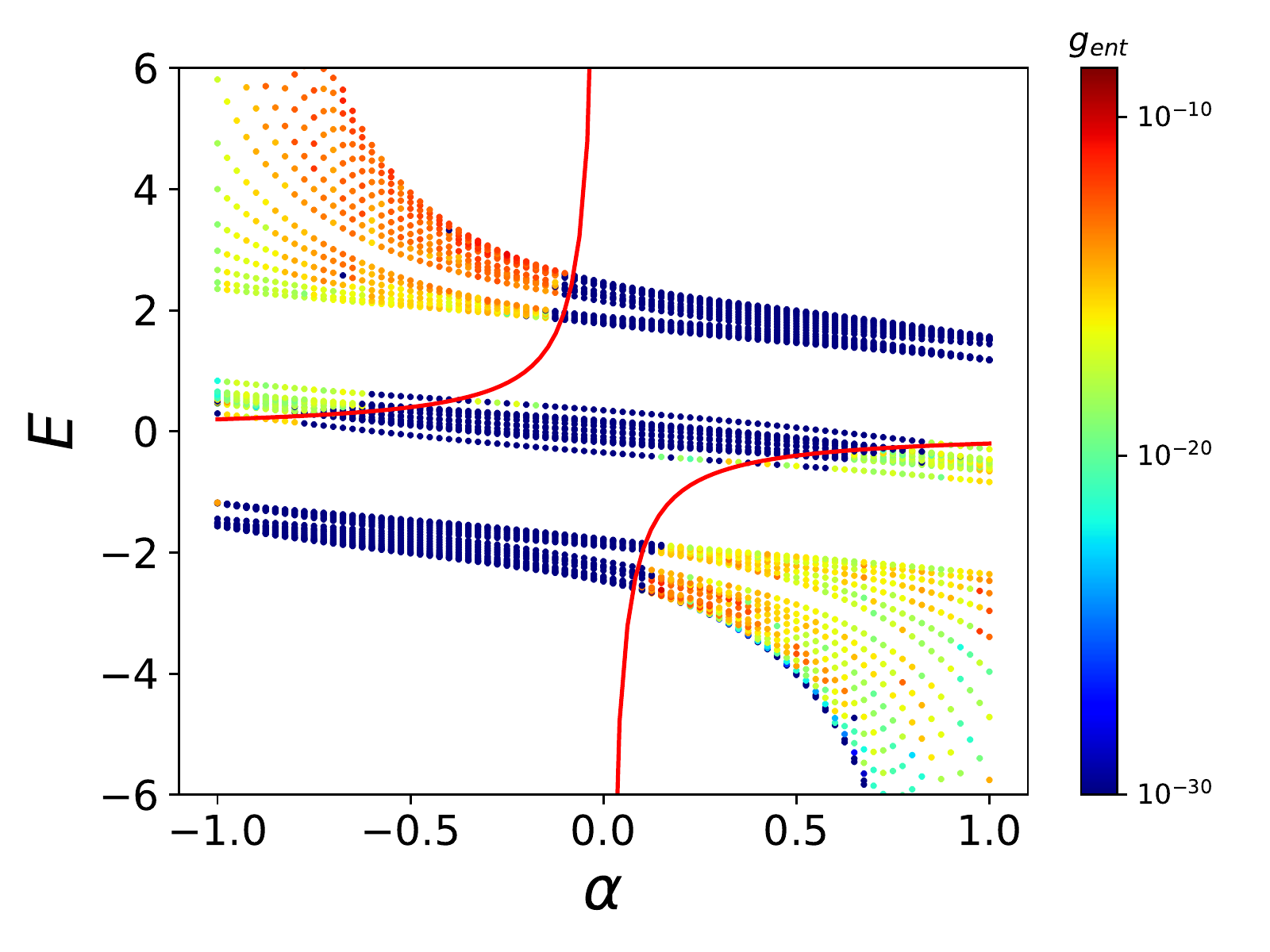}
  \end{subfigure}
  \caption{The EC ($g_{ent}$) for gAA model of $N=500$ sites as we
    change both $\lambda$ and also the number of fermions (or the
    Fermi energy). In the left panel we set $\lambda=-1.1$ and in the
    right panel $\lambda=0.9$. Red lines are the mobility edges
    according to Eq. (\ref{me}). We can see that $g_{ent}$ sharply
    determines the mobility edges between delocalized and localized
    phases.  \label{fig:g_gAA}}
\end{figure*}

\ In the previous section, we showed that by looking at the structure
of the EH matrix, different phases can be distinguished; as we go from
localized to delocalized phase, EH matrix gains more non-zero
amplitudes for far-distances hopping. Although in the pattern of the
EH matrix the difference between these phases is obviously seen (at
least in the extreme cases deep in localized and delocalized phases),
but we need a quantified measure to characterize different phases only
by a number and more importantly to identify the delocalized-localized
phase transition point. EH (which is a free fermion Hamiltonian) can
be considered as a Hamiltonian, describing a system of fermions
hopping between arbitrary sites, based on the range of the hopping
parameter. When the EH is long-range, fermions can hop between
far-distances sites and thus it is expected transportation of these
fermions to be easy. On the other hand, if EH is a short-range
Hamiltonian, hopping of the fermions would be limited to
short-distances sites and transportation becomes harder. In this
regard, conductance of a free fermion model described by the EH could
be a good candidate to distinguish long-range and short-range
Hamiltonians and consequently between delocalized and localized
phases. But, EH is not actually the Hamiltonian of a subsystem which
we then put it between two contacts and measure it conductance. So, we
will encounter conceptual difficulties if we apply the same procedure
of calculating conductance to EH. Therefore, we introduce a new
quantity, based on the conductance\cite{PhysRevLett.114.056401} in the
following way:
\begin{eqnarray}
  \nonumber G &=& H_{ent}^{-1},\\
  g_{ent} &=& G_{1N} G_{N1},
\end{eqnarray}
where $H_{ent}^{-1}$ is the inverse of the EH, $G_{1N}$ and $G_{N1}$
are respectively the $(1,N)$ and $(N,1)$ elements of the $G$
matrix. We dub $g_{ent}$ as entanglement conductance. Based on the
fact that $G$ is a symmetric matrix, $g_{ent}$ is a positive-definite
number. When we have randomness in our employed models, we have to
average over samples to obtain the $g_{ent}$. We use \textit{geometric}
averaging, since the obtained numbers ranges over large order of
magnitudes. The EC, $g_{ent}$, is plotted for AA, PRBA, and PRBM
models in Fig. \ref{fig:g_AA_PRBA_PRBM}. AA model has a short-range
Hamiltonian (in both delocalized and localized phases), on the
contrary PRBA and PRBM have a short-range Hamiltonian in localized and
a long-range one in delocalized phase. As we can see in
Fig. \ref{fig:g_AA_PRBA_PRBM}, the EC in the delocalized phase is
non-zero and it goes to zero very fast in the localized phase, and
thus determines the phase transition point exactly. In addition we
calculate the EC for gAA model (see Fig. \ref{fig:g_gAA}.)  which has
mobility edges between delocalized and localized phases. As we can see
$g_{ent}$ sharply determines the mobility edges.

\begin{figure*}
  \centering
  \begin{subfigure}{}%
    \includegraphics[width=0.32\textwidth]{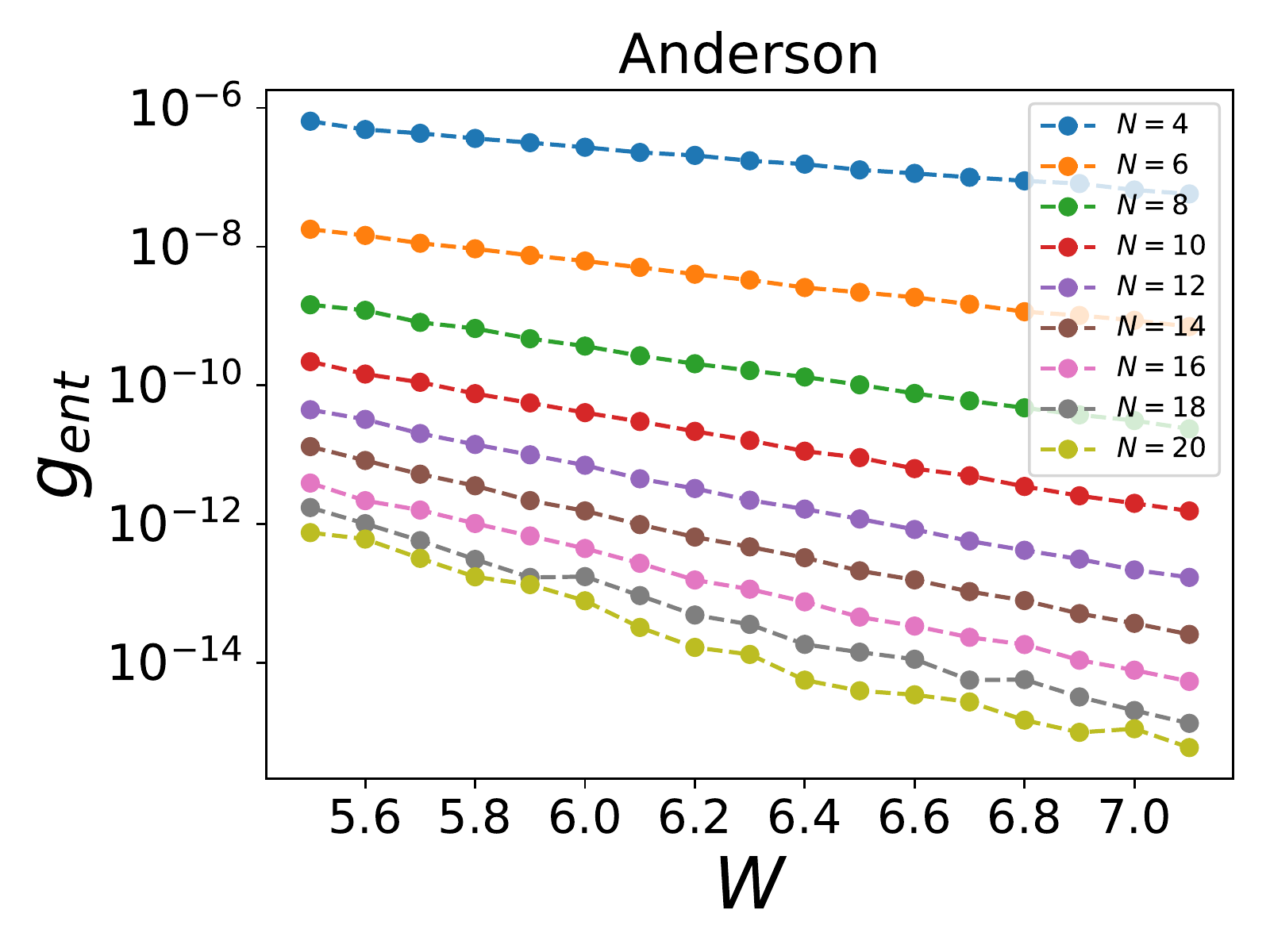}
  \end{subfigure}%
  ~%
  \begin{subfigure}{}%
  	\includegraphics[width=0.32\textwidth]{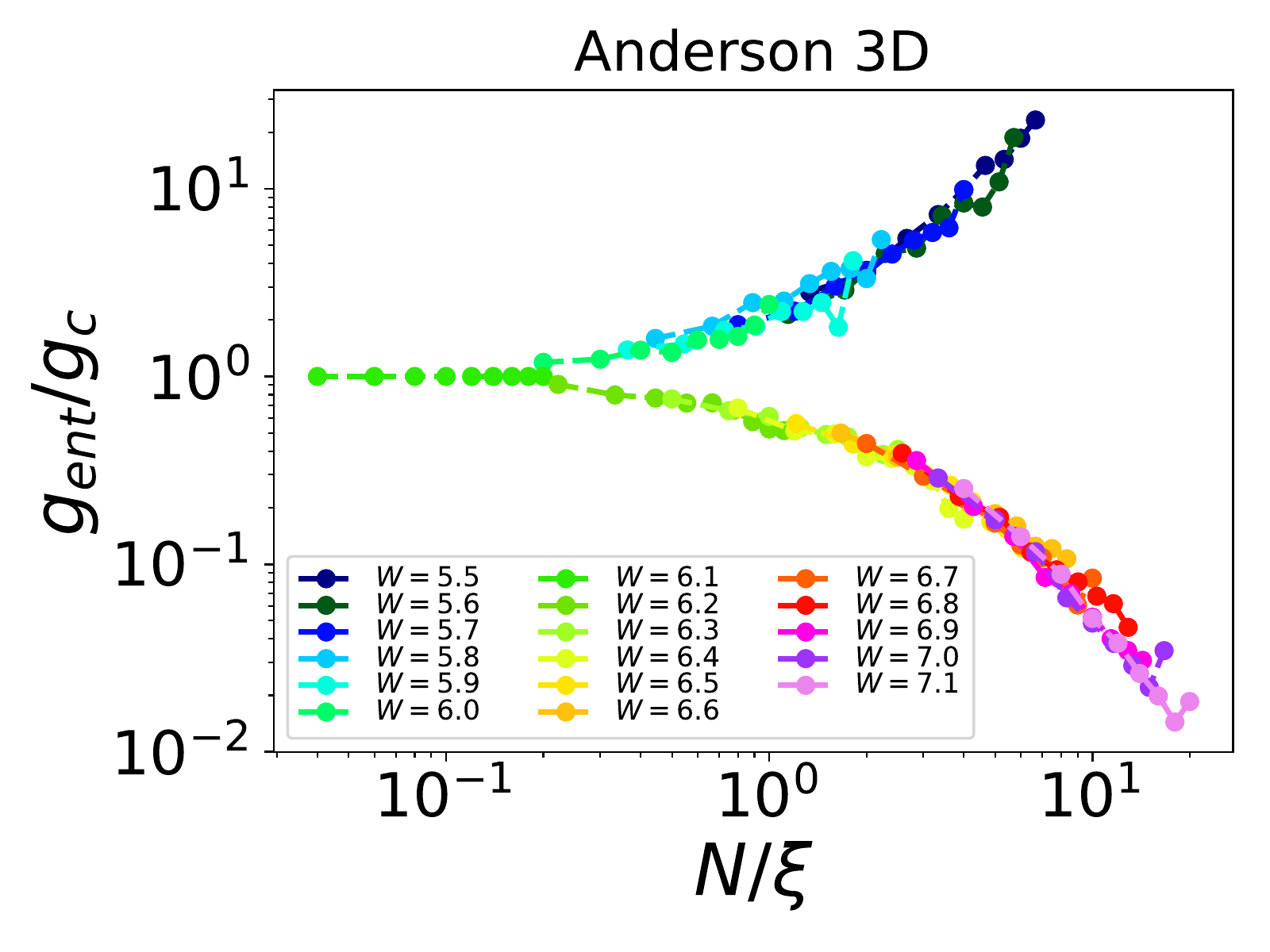}
  \end{subfigure}
  ~%
  \begin{subfigure}{}%
  	\def\big{\includegraphics[width=0.32\textwidth]{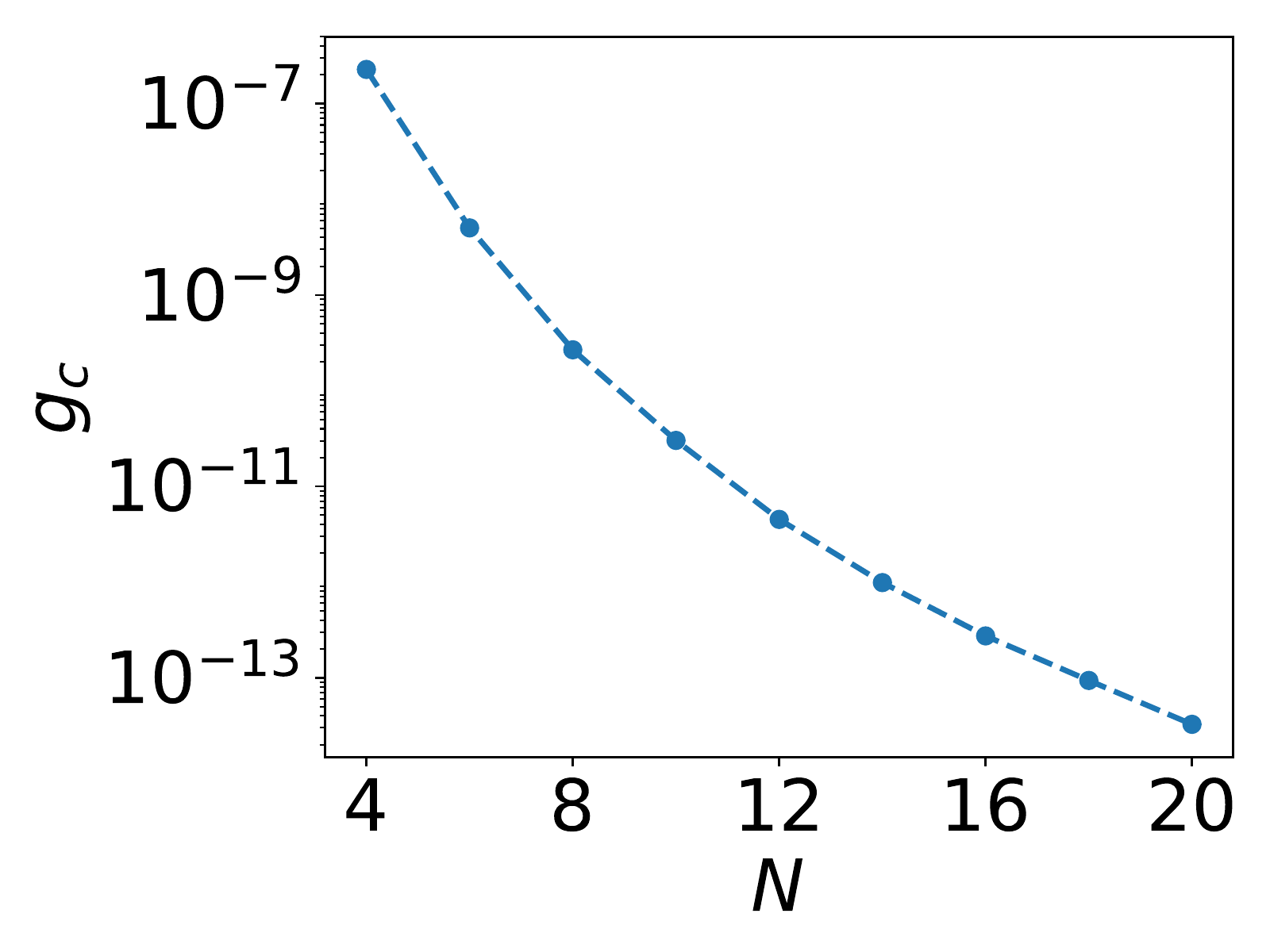}}
  	\def\little{\includegraphics[trim={10pt 15pt 10pt 10pt},clip,width=0.137\textwidth]{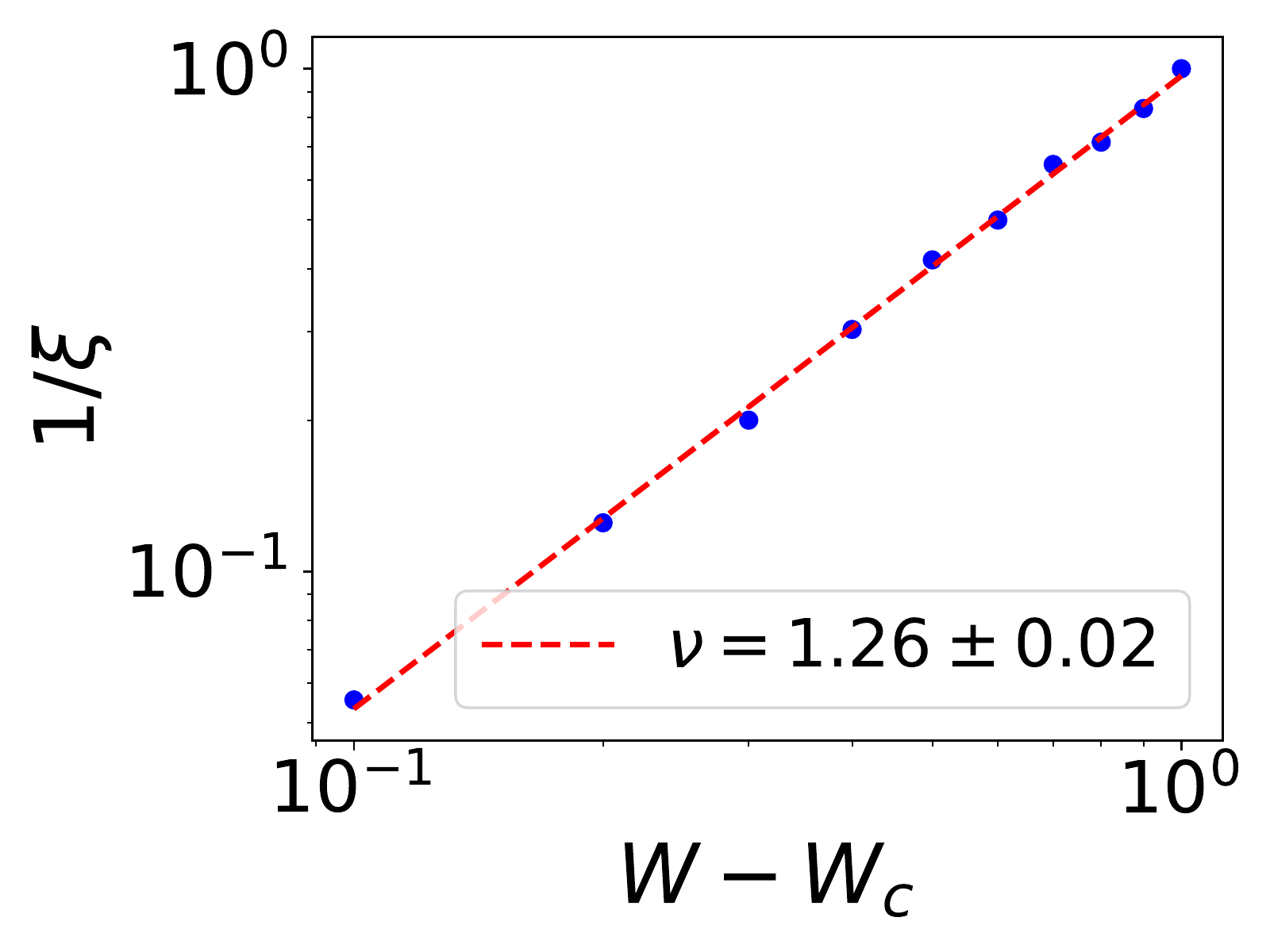}}
  	\stackinset{r}{7pt}{b}{66pt}{\little}{\big}
  \end{subfigure}
  \caption{Left panel: The EC ($g_{ent}$) for the Anderson $3D$ model
    as we vary $W$ for different system sizes. Middle panel: finite
    size scaling of EC to find the critical disorder strength. Right panel: behavior
    of $g_{ent}$ at the critical point versus system size $N$. In the
    inset plot, $1/\xi$ is plotted versus $W-W_c$ in the localized
    phase to obtain critical exponent $\nu$. The procedure of finding $W_c$ is as follows. First we choose a critical $W_c$, and plot the entanglement conductance $g_{ent}/g_c$ versus $N/\xi$, for different values of $W$. We then tune the parameter $\xi$ until we obtain two branches for EC curve (one for delocalized phase and another for localized phase). While tuning $\xi$, we do our best to have the most smooth curve. Each branch must be appeared as a continuous curve. This procedure is also applied to another choice of $W_c$. By comparing the plots of different $W_c$, we then conclude which plot is better and we choose the $W_c$. Our calculations yield to $W_c=6.1$, and $\nu \approx 1.26$. Number of sample systems
    varies between $20000$ for small $N$ and $2000$ for the largest
    $N$. \label{fig:g_Anderson_fss}}
\end{figure*}

The EC for the Anderson $3D$ model is also plotted in
Fig. \ref{fig:g_Anderson_fss}. For this model $g_{ent}$ does not go to
zero sharply in localized phase and thus it does not locate the phase
transition point; but by applying finite size scaling to the EC we are
able to calculate the critical disorder strength $W_c$ and the
corresponding critical exponent for localization length, $\nu$. Our
numerical calculations yields to $W_c=6.1$, which is consistent with
numerical results obtained before\cite{Slevin_2014}. As it
is indicated in Fig. \ref{fig:g_Anderson_fss}, $g_c$ changes with
system size $N$. So, we re-scale $g_{ent}$ by $g_c$. To obtain $W_c$
and $\nu$ we plot $g_{ent}/g_c$ versus $N/\xi$ (where $\xi$ is a
length scale for the EH, similar to the localization length of the
Hamiltonian, to show that $g_{ent}/g_c$ is scale invariant) for
different values of $W$, then we tune $W_c$ and $\xi$ to have two
branches of curves, one for delocalized and another for localized
phase. $\nu$ is given by slope of $1/\xi$ versus $W-W_c$ in log-log
scale for localized phase.

\section{Conclusion}\label{conc}
In this paper, we studied the structure of the entanglement
Hamiltonian. We showed that, independent of the Hamiltonian of the
system that can be either long-range or short-range, EH is long-range
in delocalized phase and short range in localized phase. This is due
to the fact that the EH is written in terms of single particle
correlation functions. We introduced the notion of entanglement
conductance of free fermion EH, and demonstrated that it can be served
as an order parameter for characterizing delocalizd-localized phase
transition. Entanglement conductance is a measure of how much the EH
is long-range, that is how many non-zero hopping amplitudes EH has for
far-distances sites; in one sense, it measures the amount of
entanglement in the system by looking at the structure of the
EH. Thus, to characterize the Anderson phase transition, one can look
at the amount of entanglement that increases as we go from localized
to delocalized phases; in addition and in parallel, we can say that EH
becomes long-range and consequently EH conductance increases.

\acknowledgments This work was supported by University of Mazandaran
(M. P). Part of this work was done while (M.P) was working at IASBS. We would like to thank Hossein Javan Mard for useful discussions. 

\bibliographystyle{apsrev4-1.bst}
\bibliography{References.bib}
\end{document}